\documentclass{PoS}
\newcommand{\bvec}[1]{\mbox{\boldmath $#1$}}

\title{Nuclear electric dipole moment of light nuclei in the Gaussian expansion method}

\ShortTitle{Nuclear EDM of light nuclei in the Gaussian expansion method}

\author{\speaker{Nodoka Yamanaka}\\
        iTHES Research Group, RIKEN, 2-1 Hirosawa, 351-0115 Saitama, Japan\\
        E-mail: \email{nodoka.yamanaka@riken.jp}}

\author{Emiko Hiyama\\
        RIKEN Nishina Center, RIKEN, 2-1 Hirosawa, 351-0115 Saitama, Japan\\
        E-mail: \email{hiyama@riken.jp}}

\abstract{
The nuclear electric dipole moment is a very sensitive probe of CP violation beyond the standard model, and for light nuclei, it can be evaluated accurately using few-body calculational methods. 
In this talk, we present 
the deuteron, $^3$He, $^3$H, $^6$Li, and $^9$Be electric dipole moments calculated using the Gaussian expansion method with a realistic nuclear force, and assuming the one-meson exchange model for the P, CP-odd nuclear force. 
We then give future prospects for models beyond the standard model such as the supersymmetry.
}

\FullConference{The 8th International Workshop on Chiral Dynamics, CD2015 ***\\
		29 June 2015 - 03 July 2015\\
		Pisa,Italy}

\begin{document}

\section{Introduction}

It is widely known that the standard model of particle physics has difficulty in explaining the matter abundance of the Universe.
The standard model prediction of the matter abundance is in great deficit compared with the observed matter-photon ratio.
This fact strongly suggests the existence of additional source(s) of large CP violation beyond the standard model.

To search for new CP violation, we have a very attractive experimental observable: the {\it electric dipole moment} (EDM) \cite{hereview,Bernreuther,khriplovichbook,ginges,pospelovreview,fukuyama,hewett,engel,yamanaka}.
The EDM is the permanent polarization of a system, and it is odd under parity transformation and time reversal (or equivalently CP).
The EDM can be measured in many systems such as neutrons \cite{baker}, atoms \cite{rosenberry,griffith}, molecules \cite{hudson,acme}, muons \cite{muong-2}, and many theoretical candidates beyond the standard model have been investigated so far.

In this talk, we focus on the {\it nuclear EDM}.
The nuclear EDM has many advantages.
First, it is known that the CP-odd effect is enhanced due to the nuclear many-body effect \cite{flambaum1,flambaum2,nuclearedmgem}.
It is also known that the standard model contribution is very small \cite{sushkov,smtheta,smneutronedmmckellar,donoghue,smCPVNN,czarnecki,mannel,seng}.
Finally, the experimental measurement of nuclear EDM using storage ring is in preparation \cite{storage1,storage2,storage3,storage4,storage5,bnl}.

To calculate the EDM of light nuclei, we use the Gaussian expansion method \cite{hiyama}.
In this framework, light nuclei such as the deuteron, $^3$He, $^3$H, and $^4$He can be treated ab initio, i.e. by solving directly the Schr\"{o}dinger equation of few-nucleon systems \cite{3nucleon,benchmark}.
Heavier systems such as $^6$Li and $^9$Be can also be calculated using the same method assuming the cluster approximation  \cite{clusterreview1,clusterreview2,clusterreview3}, which treats the $\alpha$ cluster as a single particle.

We now give the plan of this talk.
We first introduce the nuclear EDM and the mechanism to generate it.
We then review the Gaussian expansion method and present the result of the ab initio calculations of the EDM of the deuteron, $^3$H, and $^3$He nuclei.
In Section \ref{sec:libeedm}, we present the framework and the result of the calculation of the EDM of $^6$Li and $^9$Be in the cluster approximation.
We finally show the prospect for the search of new physics beyond the standard model and summarize our talk.

\section{Nuclear electric dipole moment and CP-odd nuclear force}

The nuclear EDM is generated in the presence of P and CP-odd nucleon level processes.
The leading contributions are the intrinsic EDM of the nucleon and the P, CP-odd nuclear force\footnote{The meson-exchange current contribution to the nucleon EDM is small \cite{liu,stetcu}, although that to the magnetic moment is not small for some nuclei \cite{pastore}.}.
The nucleon EDM contribution to the nuclear EDM is given by
\begin{equation}
d_A^{\rm (Nedm)} 
=
\sum_{i=1}^A
d_i \langle \,A\, |\, \sigma_{iz} \, |\, A\, \rangle
\equiv 
\langle \sigma_p \rangle_A \, d_p + \langle \sigma_n \rangle_A \, d_n
,
\end{equation}
with $A$ the number of nucleons of the nucleus and $|\, A\, \rangle$ the nuclear wave function polarized in the $z$-direction. 
The proton and the neutron EDMs are given by $d_p$ and $d_n$, respectively.
The nuclear EDM generated by the P, CP-odd nuclear force is
\begin{equation}
d_{A}^{\rm (pol)} 
=
\sum_{i=1}^{A} \frac{e}{2} 
\langle \, \tilde A \, |\, (1+\tau_i^z ) \, {\cal R}_{iz} \, | \, \tilde A \, \rangle
,
\end{equation}
with $\tau^z_i$ the isospin Pauli matrix, and ${\cal R}_{iz}$ the coordinate of the nucleons in the nuclear center of mass frame.
Here $|\, \tilde A\, \rangle$ is the nuclear wave function polarized in the $z$-axis, including a small component of opposite parity states.
It is to be noted that the permanent polarization of the nucleus is realized only if there is a mixing of opposite parity states.

To generate such parity mixing and the nuclear polarization, the P, CP-odd nuclear force is necessary.
In this work, we model the CP-odd nuclear force by the CP-odd one-pion exchange $N-N$ interaction \cite{liu,pvcpvhamiltonian1,pvcpvhamiltonian2,pvcpvhamiltonian3}.
The CP-odd Hamiltonian used in our work is given by
\begin{eqnarray}
H_{P\hspace{-.5em}/\, T\hspace{-.5em}/\, } 
& = &
\frac{1}{2m_N} \Bigl[
\ \ \ \ \, \bar{G}_{\pi}^{(0)}\, \bvec{\tau}_{1}\cdot \bvec{\tau}_{2}\,\bvec{\sigma}_{-}\cdot \bvec{\nabla}
\mathcal{Y}_{\pi}(r)
\nonumber \\
&& \hspace{3em} 
+ \bar{G}_{\pi}^{(1)}\, ( \tau_1^{z}\,\bvec{\sigma}_1 -\tau_2^{z}\,\bvec{\sigma}_2 ) \cdot \bvec{ \nabla}\,\mathcal{Y}_{\pi}(r)
\nonumber \\
&& \hspace{3em}
+\bar{G}_{\pi}^{(2)}\, (3\tau_{1}^{z}\tau_{2}^{z}-\bvec{\tau}_{1}\cdot \bvec{\tau}_{2})\,\bvec{\sigma}_{-} \cdot \bvec{\nabla} \mathcal{Y}_{\pi}(r)
\ \ \Bigr]
,
\label{eq:CPVhamiltonian}
\end{eqnarray}
where $\bar{G}_\pi^{(i)} \equiv g_{\pi NN} \bar g_{\pi NN}^{(i)}$ ($i=0,1,2$) is the CP-odd coupling constant, generated by new physics beyond the standard model.
We consider these unknown CP-odd couplings as small. 
The radial part of the CP-odd Hamiltonian is made of the derivative of the Yukawa function ${\cal Y}_\pi (r) \equiv  \frac{e^{-m_\pi r }}{4 \pi r}$.
Due to the small CP violation, it is sufficient to consider the nuclear EDM in the linear approximation:
\begin{equation}
d_A^{\rm (pol)} 
=
\sum_{i} a_{A,\pi}^{(i)} \bar G_\pi^{(i)} 
.
\end{equation}
The linear coefficients $a_{A,\pi}^{(i)}$ depend only on the nuclear structure, and can be calculated using the Gaussian expansion method \cite{nuclearedmgem}.

\section{Gaussian expansion method}

We now very briefly introduce the Gaussian expansion method \cite{hiyama}.
The general wave function of the three-body system is given as a superposition of Gaussian functions:
\begin{eqnarray}
\Psi_{JM, TT_z}
&=&
\sum_{c}
\sum_{T}
\sum_{\Sigma}
\sum_{\Lambda}
\sum_{nl, NL}
C^{(c)}_{nl,NL,\Sigma,T}
{\cal S}_\alpha
\Big[ 
[ \phi^{(c)}_{nl}({\bf r}_c)
\psi^{(c)}_{NL}({\bf R}_c)]_\Lambda \, 
\chi_\Sigma
\Big]_{J,M}
\eta_{T , T_z}
. \ \ \ 
\label{eq:he7lwf}
\end{eqnarray}
Here the operator $\cal{S_\alpha}$ denotes the symmetrization (antisymmetrization) between the identical bosons (fermions).
The spin and isospin functions are denoted by $\chi_\Sigma$ and $\eta_{T , T_z}$, respectively.

The radial components of the wave function $\phi_{nlm}({\bf r})$, $\psi_{NLM}({\bf R})$ are given by
\begin{eqnarray}
\phi_{nlm}({\bf r})
&=&
r^l \, e^{-(r/r_n)^2}
Y_{lm}({\widehat {\bf r}})  \;  ,
\nonumber \\
\psi_{NLM}({\bf R})
&=&
R^L \, e^{-(R/R_N)^2}
Y_{LM}({\widehat {\bf R}})  \;  ,
\end{eqnarray}
where the Gaussian range parameters obey the following geometric progression:
\begin{eqnarray}
      r_n
      &=&
      r_1 a^{n-1} \qquad \enspace
      (n=1 - n_{\rm max}) \; ,
\nonumber\\
      R_N
      &=&
      R_1 A^{N-1} \quad
     (N \! =1 - N_{\rm max}).
\end{eqnarray}
In our calculation, we take the angular momentum space $l, L, \Lambda \leq 2$, which shows good convergence of the calculation.

\section{Test of ab initio calculations ($^2$H, $^3$He, and $^3$H nuclei)}

We have calculated the EDM of the deuteron, $^3$H, and $^3$He nuclei to test the ab initio calculation using the Gaussian expansion method.
As the Hamiltonian, we have used the Argonne $v18$ interaction \cite{av18} and the one-pion exchange CP-odd potential of Eq. (\ref{eq:CPVhamiltonian}).
The three-body force has not been considered.
The results are shown in Table \ref{table:abinitionuclearedm}.

For the deuteron, the result is quite in good agreement with the values of Refs. \cite{liu,afnan,song}.
We also note that the coefficients of the intrinsic nucleon EDM contribution for the deuteron are smaller than one, due to the $d$-wave component.
For the EDMs of $^3$He and $^3$H, our result agrees again with the chiral effective field theory calculations of Refs. \cite{bsaisou,bsaisou2}.
The agreement with previous works certifies that the Gaussian expansion method works well as an ab initio method.

\begin{table}
\begin{center}
\caption{
Results of the ab initio calculation of the EDM of the deuteron, $^3$H, and $^3$He nuclei.
The unit of the linear coefficients of the CP-odd $N-N$ coupling $a_\pi^{(i)}$ ($i=0,1,2$) is $10^{-2} e$ fm.
The symbol $-$ denotes that the result cancels in our setup.
The coefficient of the neutron EDM calculated in the chiral analysis \cite{crewther} was also added for comparison.
}
\begin{tabular}{l|cc|ccc}
\hline
\hline
&$\langle \sigma_p \rangle$ & $\langle \sigma_n \rangle$ &$a_\pi^{(0)}$ & $a_\pi^{(1)}$ & $a_\pi^{(2)}$  
\\ 
\hline
\ $n$  & 0 & 1 & 1 & $-$ & $-1 $     
\\
$^{2}$H  & 0.914 & 0.914 & $-$ & $1.45 $ & $-$ 
\\
$^{3}$He  & -0.04 & 0.89 & $0.59$ & 1.08 & 1.68 
\\
$^{3}$H & 0.88 & -0.05 & $-0.59$ & 1.08 & -1.70  
\\
\hline
\hline
\end{tabular}

\label{table:abinitionuclearedm}
\end{center}
\end{table}

\section{Results of the calculation in the cluster approximation ($^6$Li and $^9$Be nuclei)\label{sec:libeedm}}

In the cluster approximation, $^6$Li and $^9$Be are considered as $\alpha +n+p$ \cite{hasegawa,Kukulin} and $\alpha +\alpha +n$ \cite{revai,orlowski,arai} three-body systems, respectively.
The CP-even effective $\alpha - N$ and $\alpha - \alpha$ interactions are obtained from the fit which reproduces the low energy scattering phase shift of the corresponding systems \cite{kanada, hasegawa}.
We use the Argonne $v8'$ interaction \cite{av18} for the $N-N$ interaction which is required for the calculation of the $^6$Li nucleus.

The effect of the antisymmetrization for the $N-\alpha$ and $\alpha - \alpha$ subsystems is approximated by the orthogonality condition model (OCM) \cite{ocm1,ocm2,ocm3}.
It is possible to project out the forbidden states by including the following term in the Hamiltonian \cite{Kukulin}
\begin{equation}
V_{\rm Pauli}=\lim_{\lambda \rightarrow \infty}\sum_{f} 
\lambda \, |\phi_f(\mbox{\boldmath r}_{\alpha x}) \rangle  \langle \phi_f ( \mbox{\boldmath r'}_{\alpha x} ) |.
\end{equation}
In our calculation, $\lambda$ was taken to $10^4$ MeV.

\begin{figure}[htb]
\begin{center}
\includegraphics[width=9cm]{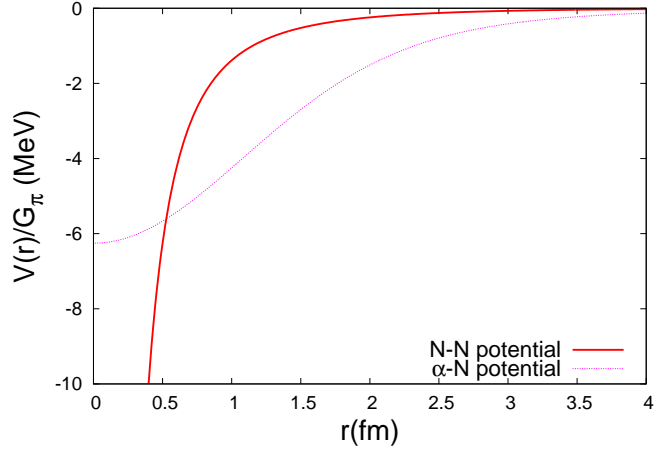}
\caption{\label{fig:folding}
The folded CP-odd $\alpha - N$ potential. 
The CP-odd $N - N$ potential is also shown for comparison.
We have factored out the unknown coupling constant $\bar G^{(1)}_\pi$.
}
\end{center}
\end{figure}

The CP-odd $\alpha -N$ interaction is obtained by the folding
\begin{equation}
V_{\alpha-N} ({\mathbf r})
=
\frac{\bar{G}_{\pi}^{(1)}}{2 m_N} 
\int d^3{\mathbf r'} \rho_\alpha ({\mathbf r'}) \, 
 \bvec{ \nabla}\,\mathcal{Y}_{\pi} ({\mathbf r}-{\mathbf r'})
,
\end{equation}
where the nucleon number density of the $\alpha$ cluster is $\rho_\alpha (r) = \frac{4}{b^3 \pi^{3/2}} e^{-r^2 /b^2}$ with $b=$ 1.358 fm.
The shape of the folding potential is shown in Fig. \ref{fig:folding}.
The isoscalar and isotensor CP-odd nuclear forces cancel after the folding.
Moreover, all CP-odd nuclear forces cancel for the $\alpha - \alpha$ system.

\begin{table}
\begin{center}
\caption{
Results of the EDM calculations of the cluster approximation.
The unit of the linear coefficients of the CP-odd $N-N$ coupling $a_\pi^{(i)}$ ($i=0,1,2$) is $10^{-2} e$ fm.
The symbol $-$ denotes that the result cancels in our setup.
}
\begin{tabular}{l|cc|ccc}
\hline
\hline
  &$\langle \sigma_p \rangle$ & $\langle \sigma_n \rangle$ &$a_\pi^{(0)}$ & $a_\pi^{(1)}$ & $a_\pi^{(2)}$ 
\\ 
\hline
$^{6}$Li & 0.88 & 0.88 & $-$ & 2.8 & $-$ 
\\
$^{9}$Be & $-$ & 0.75 & $-$ & $1.7$ & $-$ 
\\
\hline
\hline
\end{tabular}

\label{table:clusternuclearedm}
\end{center}
\end{table}

The result of the calculation of the EDM of the $^6$Li and $^9$Be nuclei in the cluster approximation is shown in Table \ref{table:clusternuclearedm}.
The EDM of $^6$Li is remarkably sensitive to the isovector CP-odd one-pion exchange nuclear force.
It is made of two comparable contributions, namely the EDM of the deuteron subcluster  and the polarization due to the CP-odd $\alpha - N$ interaction.
The $^6$Li EDM is two times more sensitive than the deuteron EDM.

For $^9$Be, its sensitivity to the isovector CP-odd nuclear force is comparable to that of the deuteron.
This time, the EDM is only due to the CP-odd $\alpha -N$ interaction.
Its value coincides with the same polarization effect in $^6$Li.

\section{Impact to new physics beyond the standard model}

We now discuss the prospects for the search of new physics beyond the standard model.
Let us first discuss the sensitivity to supersymmetric models.
The supersymmetric models contribute to the fermion EDM at the one-loop level, with supersymmetric particles in the intermediate state \cite{pospelovreview,ellis}.
If the sensitivity of $O(10^{-29})e$ cm is reached in the planned EDM experiment using storage rings, we can probe the supersymmetry breaking mass scale at the level of 10 TeV.

In models generating a 4-quark interaction with the exchange of new bosons, the mass scale of 10 TeV to PeV can be probed.
This is the case of the Left-right symmetric model \cite{Dekens:2014jka} or leptoquark model \cite{herczeg}.

Other interesting targets are models contributing to the Barr-Zee type diagram.
As examples, we have the Higgs doublet models \cite{barr-zee,gunion,chang,hestrange,hayashi,bowser-chao,abe,jung}, or the R-parity violating supersymmetric models \cite{yamanaka,rpv1,rpv2,rpv3,rpv4}.
The prospective experimental sensitivity can probe the new physics  mass scale of $\sqrt{Y_q Y_Q}$ PeV, with $Y_q$ and $Y_Q$ the couplings between the exchanged new boson and fermions.

\section{Summary}

In this talk, we have presented the results of the calculation of the EDM of $^2$H, $^3$He, $^3$H, $^6$Li, and $^9$Be using the Gaussian expansion method.
The results show that $^6$Li is sensitive to the CP-odd nuclear force due to the constructive interference between the CP-odd $\alpha - N$ interaction and the deuteron EDM.
The planned experiment using storage ring has a sensitivity of $O(10^{-29})e$ cm, and we expect it to unveil the  CP violation of new physics beyond the TeV scale.

\acknowledgments{
The authors thank Saori Pastore for useful discussions and comments.
This work is supported by RIKEN iTHES Project.
}

\end{document}